\documentstyle[fleqn]{article}
\newcommand{\be}{\begin{equation}}
\newcommand{\ee}{\end{equation}}
\newcommand{\bi}[1]{\vspace{-3mm}  \bibitem{#1}}

\begin{document}

\begin{center}
{\Large \bf Analog of Lie Algebra and Lie Group 
 for Quantum Non-Hamiltonian Systems. }\\
\vspace{4mm}

Vasily E.Tarasov, \\
Nuclear Physics Institute, Moscow State University, \\
119899 Moscow, RUSSIA\\
E-mail: TARASOV@THEORY.NPI.MSU.SU \\
 
\end{center}

\begin{abstract}
\begin{center}
In order to describe non-Hamiltonian (dissipative) systems in quantum theory
we need to use non-Lie algebra such that commutators of this algebra
generate Lie subalgebra.
It was shown that classical connection between analytic group
(Lie group) and Lie algebra, proved by Lie theorems, 
exists between analytic loop, 
commutant of which is associative subloop (group), 
and commutant Lie algebra (an algebra
commutant of which is Lie subalgebra).
\end{center}
\end{abstract}



\section{Introduction.}

Quantum mechanics of Hamiltonian (non-dissipative) systems
uses well known pair: Lie algebra and analytic group (Lie group).
Elements of this pair are connected by Lie theorems.
In order to describe non-Hamiltonian (dissipative) systems in quantum
theory \cite{Tarydfiz,Tartmf1,Tarictp,Tar3} we need to use
non-Lie algebra and analytic quasigroup (loop).
This new pair consists of commutant Lie algebra (Valya algebra)
and analytic commutant associative loop (Valya loop).
A commutant Lie algebra (Valya algebra) is 
an algebra the commutators of which generate Lie subalgebra
(an algebra such that commutant is a Lie subalgebra).
Commutant Lie algebra can be defined by
the following conditions $x^{2}=0 \ ; J(xy, zq, pl) =0 \ , $
where $ J(x,y,z) = (xy)z + (yz)x + (zx)y $ -- Jacobian of elements $x,y,z$.
Valya loop is a non-associative loop such that the commutant of this loop 
is associative subloop (group). 
It is known that a commutant of algebra is a subspace,
which is generated by all commutators of the algebra.
We proved that a tangent algebra of Valya loop is
a commutant Lie algebra (Valya algebra) -- it is
analog of Lie theorem.
It was shown that generalized Heisenberg-Weyl algebra, suggested 
by the author \cite{Tarydfiz,Tartmf1,Tarictp} to describe quantum
non-Hamiltonian (dissipative) systems, is a commutant Lie algebra.
As the other example of commutant Lie algebra, it was considered
a generalized Poisson algebra \cite{Tar3} for differential 1-forms.


\section{ Generalized Heisenberg-Weyl algebra.}

{\bf 2.1. Definition of Generalized Heisenberg-Weyl Algebra.}

Usual  Heisenberg-Weyl algebra  $W_N$ is defined by 
basic elements $\{Q_i, P_i, I \} $ which are  satisfied
the following commutation relations 
\be
\label{gcr11}
[ Q_i, Q_j ] = [ P_i, P_j ] =  [ Q_i, I ] = [ P_i, I ] = 0 \ ,
\qquad [ Q_i, P_j ] = \imath  \delta_{ij} I  
\ee
Basis of generalized  Heisenberg-Weyl algebra
$W^*_N$ consists of $2N+2$
basic elements $\{Q_i, P_i, W, I \} $ which are satisfied
the relations (\ref{gcr11}) and commutation relations
\cite{Tarydfiz,Tartmf1,Tarictp,Tar3}:
\be
\label{gcr12}
 [ W, P_i ] = \imath  F_i (Q,P)  \ ,
\qquad [ W, Q_i ] = [W, W] = [ W, I ] = 0 
\ee
In the relation (\ref{gcr12}) elements $F_i (Q,P)$ are functions
(for example, polynomial) of basic elements ${Q_i, P_i }$.
In the simplest case which is interesting in application, the relation
(\ref{gcr12}) has the form
$ [ W, P_i ] = \imath  \gamma_{ij} P_j \ , $ 
where $ \gamma_{ij} $ --  numbers.
Generalized Heisenberg-Weyl algebra for this case we denote by
$LW^*_N$.
Let us note that physical meaning of elements  $F_i (Q,P)$ are the
following:   $F_i (Q,P)$ is  dissipative force (friction) which
is acts on the system.

{\bf 2.2. Identities for Jacobians.} 

Commutation relations (\ref{gcr12}) lead to 
the fact that some Jacobians of basics elements
are not equal to zero
\cite{Tarydfiz,Tartmf2,Tar3}:
\be
\label{jac3}
J[P_i, P_j, W ] = \imath ([F_i, P_j] - [F_j, P_i]) \ ,
\qquad J[Q_i, P_j, W ] =  - \imath  [F_j, Q_i] \ ,
\ee
where $ \  J[x,y,z] =  \ [[x, y], z] + [[y, z], x] + [[z, y], x]  $
is Jacobian of elements $x,y,z$.
The fact that Jacobians (\ref{jac3}) are not equal to zero 
is a main distinction of non-Hamiltonian systems and
it is characteristic property
of suggested generalization of Heisenberg-Weyl algebra.
In the consequence of relations (\ref{jac3}) generalized 
Heisenberg-Weyl algebra $W^*_N$ is non-Lie algebra.

{\bf 2.3. General Element of Generalized Heisenberg-Weyl Algebra.} 

Let us consider general element $Z$ of generalized Heisenberg-Weyl algebra 
$W^*_N$ 
\[ Z = sI + x_i Q_i + y_i P_i + t W \ , \]
where $s, x_i, y_i, t$  -- numbers.
Commutator of general elements $Z_1$ and $Z_2$ has the form
\[ [Z_1,Z_2] = \imath s_3 I + \imath l_i F_i (Q,P) \ , 
\ \ where \ \ s_3 = x^1_i y^2_i - x^2_i y^1_i  
\quad l_i = t^1 y^2_i - t^2 y^1_i \ . \]  
Jacobian of general elements $Z_1$, $Z_2$ and $Z_3$  has the form
\[ J[Z_1,Z_2,Z_3] = i s_{ij} [Q_i,F_j]  + 
i t_{ij}  ([F_i, P_j] - [F_j, P_i])  \ , \]
\[where \ \ s_{ij} = (x^1_i y^2_j - x^2_i y^1_j) t^3 +
 (x^1_i y^3_j - x^3_i y^1_j) t^2 + (x^2_i y^3_j - x^3_i y^2_j) t^1 \]
\[t_{ij} = 2 (y^2_i y^3_jt^1 + y^3_i y^1_j t^2 + y^1_i y^2_j t^3) \quad . \]

In the simplest case with $F_i (Q,P) = \gamma_{ij} P_j $
the commutator of general elements $Z_1$ and $Z_2$ has the form
\[ [Z_1,Z_2] = \imath Z_3  \ , \qquad
where \ \ s_3 = x^1_i y^2_i - x^2_i y^1_i \ \  x^3_i =0 
\ \ y^3_i = t^1 y^2_i - t^2 y^1_i \ \  t^3 = 0 \ , \]
and Jacobian of elements $Z_1$, $Z_2$ and $Z_3$  is
\[ J[Z_1,Z_2,Z_3] = Z_4 \ , \qquad
where \ \ s_4 = - s_{ij} \gamma_{ij} \qquad x^4_i=y^4_i=t^4=0 \ \ . \]


\section{  Commutant Lie Algebra.}

{\bf 3.1.} Let us introduce 

DEFINITION 1. An algebra B will be called {\bf Valya algebra},
if multiplicative operation is satisfied the following conditions

1) Anti-symmetric identity: $ \quad xx=0 \ ; $

2) Soft Jacobi identity: $\quad J(xy, zp, ql) =0 \ , $

where $ \ J(x,y,z) = (xy)z + (yz)x + (zx)y \ .$

It is easy to see that Lie algebra, which is defined by the conditions
$x^2=0, \ J(x,y,z)=0 $, is Valya algebra. A binary Lie algebra, 
the basic operation of which satisfies the conditions 
$ x^2=0, \ J(x,y,xy) =0$, is Valya algebra.

Let $A$ be a non-associative algebra. Let us introduce 
a binary operation: $[x,y] = xy - yx$ which is called commutator. 
As the result we derive anti-commutative algebra $A^{(-)}$,
which is associated with the algebra $A$. For this
algebra anti-symmetric and soft Jacobi conditions
are realized in the form

1) Anti-symmetric identity: $ \quad [x,y]=0 \ ; $

2) Soft Jacobi identity: $ \quad J[[x,y],[z,p],[q,l]]=0 \ . $

Let us denote by $g=A^{(-)}$  an anti-commutative algebra $A^{(-)}$,
which is associated with non-associative algebra A.
It is known the following  

DEFINITION 2.
A {\bf commutant} of the algebra $g$ is a subspace $[g,g]=g^{\prime}$,
which is generated by all commutators  $[x,y]$, where $x,y \in g$.

DEFINITION 3.
A {\bf divisible commutants} $g^{(k)} \ k=0,1,2,... $ of the algebra $g$ 
are defined by the rule $ g^{(0)}=g \quad g^{(k+1)}=(g^{(k)})^{\prime} $

It is easy to prove the following 

PROPOSITION \\
1. A commutant of algebra $g$ is ideal, but a commutant $g^{(2)}$
of commutant  $g^{(1)}$ is not ideal of this algebra $g$.\\ 
2. A divisible commutant $g^{(k+1)}$  is ideal of the algebra
$g^{(k)}$ , but is not ideal of algebra $g^{(l)}$ , where $l < k$.\\
3. If Jacobi identity is satisfied then 
a divisible commutant $g^{(k+1)}$ is ideal of the algebra $g^{(l)}$ , 
where $l < k$.

Let us introduce a new definition

DEFINITION 4.
An algebra  $g=A^{(-)}$ will be called a {\bf commutant Lie algebra}, 
if the commutators of this algebra generate subalgebra
which is Lie algebra, or if the commutant $g^{\prime}$ of the algebra $g$ 
is Lie subalgebra.

It is easy to prove the following

PROPOSITION.
A commutant Lie algebra is Valya algebra.

PROOF: \\
Let us consider elements $z_k $ of commutant Lie algebra 
$A^{(-)}$, where $k=1,2,...,6$. Using the definition of
Lie algebra and definition of commutant Lie algebra the following
conditions are satisfied

1) Anti-symmetric identity:  $ \quad [z_k,z_l]=0 \ ;$

2) Soft Jacobi identity: $ \quad J[[z_1,z_2],[z_3,z_4],[z_5,z_6]]=0 $ 

{\bf 3.2.} Let us introduce the following new definition

DEFINITION 5.
An algebra A will be called a 
{\bf commutant associative algebra}, if 
associator of commutators is equal to zero, that is 
the basic operation satisfies the following conditions 
\be
\label{comas}
 (xy-yx,zp-pz,ql-lq)=0 \ , 
\ee
\[ or  \ \  (xy,zp,ql)+(xy,pz,lq)+(yx,zp,lq)+(yx,pz,ql) - \] 
\[ -(yx,zp,ql)-(xy,pz,ql)-(xy,zp,lq)-(yx,pz,lq)=0 \ , \]
where $(x,y,z) = (xy)z -x(yz)$ -- associator of algebra elements.

THEOREM. \\
Let $A$ be a commutant associative algebra. 
An anti-commutative algebra $A^{(-)}$ of this commutant 
associative algebra $A$ is a commutant Lie algebra.

PROOF:\\
Let us use the connection between Jacobian and associator in the form
\[ J[x,y,z] = (x,y,z) +(y,z,x) +(z,x,y)-(z,y,x)-(y,x,z)-(x,z,y) \]
If we consider the Jacobian of commutators
\[ J[[x,y],[z,p],[q,l]]=J[xy-yx,zp-pz,ql-lq] \]
and definition condition (\ref{comas}) of commutant associative algebra
then we derive soft Jacobi identity  and anti-symmetric identity
for the algebra $A^{(-)}$.

{\bf 3.3.} It is easy to prove the following

PROPOSITION.\\
A generalized Heisenberg-Weyl algebra  $W^*_N$ is a commutant Lie algebra 
(Valya algebra), which is non-Lie algebra. 

PROPOSITION.\\
A generalized Heisenberg-Weyl algebra $LW^*_N$, which is defined by
the relations (\ref{gcr11}-\ref{gcr12}) with 
$ F_i (Q,P) = \gamma_{ij} P_j $, satisfies not only
soft Jacobi identity but the following conditions are satisfied 
\be
1) \ J[[Z_1,Z_2],[Z_3,Z_4],Z_5]=0 \ ; \
2) \ [J[Z_1,Z_2,Z_3],Z_4] =0  \ ; \
3) \ [[Z_1,Z_2],[Z_3,Z_4]] =0  
\ee
Note that first identity and soft Jacobi identity are
the consequence of third condition. It is easy to prove the following

PROPOSITIONS.\\
1. Heisenberg-Weyl algebra $W_N$ is an ideal of generalized
 Heisenberg-Weyl algebra $W^*_N$.\\
2. A commutant of generalized Heisenberg-Weyl algebra $W^*_N$ 
is a subalgebra of Heisenberg-Weyl algebra  $W_N$. \\
3. Maximal ideal of generalized  Heisenberg-Weyl algebra $W^*_N$, 
which is Lie algebra, is  Heisenberg-Weyl algebra $W_N$. \\
4. Annulator of generalized Heisenberg-Weyl algebra  $W^*_N$ is an
annulator of Heisenberg-Weyl algebra  $W_N$  \{ Z : Z=sI \}.


\section{ Loops with Associative Commutants.}

{\bf 4.1. Loop with Inverse Elements and Loop Commutant.} 

Analytic loop -- non-associative generalization of 
analytic group (Lie group) -- first consider by A.I. Malcev \cite{a1},
see also \cite{a2,a3,a4,a5}.

Let $G$ be a{\bf loop with inverse elements} \cite{a5}, 
that is a set $G$ with binary operation $( \circ )$ such that the following
conditions are satisfied

1) There exists a fixed element $e \in G$ such that 
$x \circ e = e \circ x = x$ for all $x$ in $G$;

2) For each element $a \in G$ there exists unique inverse
element $a^{-1}$ in $G$ such that:
$ a \circ a^{-1}=a^{-1} \circ a=e $

3) For all elements $a, b \in G$ the following identity are satisfied:
$ (b \circ a) \circ a^{-1} = a^{-1} \circ (a \circ b)= b $

Note that all elements $a,b$ in this loop $G$ 
satisfy the following important identity
\cite{a5}:
$$(a \circ b)^{-1}= b^{-1} \circ a^{-1} $$

It is known that a commutator of the loop with inverse element can be defined.
A {\bf commutator} of the elements $x,y \in G$ is a element $z$ 
in $G$ such that 
$$ z = [x,y]= (x \circ  y) \circ (y \circ x)^{-1}
= (x \circ y)  \circ (x^{-1} \circ y^{-1} ) $$

DEFINITION 6.
A {\bf commutant of loop} $G$ is a set $G^{(-)}$ 
of elements $z$ in $G$,
which can be represented in the form $z=z_1z_2...z_m$ , where
$z_i$ are commutators of elements $x_i,y_i \in G$.

{\bf 4.2. Tangent Algebra of Analytic Loop.}

Following \cite{a1} we shall call an {\bf analytic loop} 
an analytic manifold, 
such that the binary operation on the manifold satisfies 
loop structure conditions and
is analytic operation \cite{a1}.

A {\bf tangent algebra} $g$ of local analytic loop G is 
a tangent space $T_e(G)$ together with binary and ternary operations
$[\ ,\ ]$ and $< \ , \ , \ >$ which are defined in the following  
form \cite{a3}. 
Let $\alpha(t), \ \beta(t), \ \gamma(t)$ -- smooth curves in a loop G, 
going through the point $e$   $\alpha(0)= \beta(0)=\gamma(0)=e$
and having tangent vectors
 $\alpha^{\prime}(0)=\xi, \ \beta^{\prime}(t)=\eta, 
 \ \gamma^{\prime}(t)=\zeta$. Then
\be
(\beta(t) \alpha(t)) \backslash (\alpha(t) \beta(t)) =t^2 [\xi,\eta] + o(t^2)
\ee
\be
(\alpha(t) (\beta(t) \gamma(t))) \backslash 
((\alpha(t) \beta(t)) \gamma(t)) = t^3 <\xi,\eta,\zeta>+o(t^3)
\ee

Using local coordinate on analytic loop $G$
in the neighborhood of the point $e$ the production $z=x\circ y$
can be expanded in the Teylor series $z_i=\mu_i(x,y)$ 
in the form
\be
\label{teylor}
\mu_i(x,y)= x_i+y_i+a^i_{jk}x_jy_j +b^i_{jkl} x_jx_ky_l 
+c^i_{jkl} x_jy_ky_l+... 
\ee
Binary and ternary operations have the form
\be
\label{tang}
[\xi,\eta]_i = u^i_{jk} \xi^j \eta^k \quad <\xi,\eta,\zeta>_i =
v^i_{jkl} \xi^j \eta^k \zeta^l
\ee
\be
where \ \ u^i_{jk}=a^i_{jk}-a^i_{kj} \ , \quad 
v^i_{jkl}=2b^i_{jkl}-2c^i_{jkl}+\frac{1}{4} u^i_{ml}v^m_{jk}
-\frac{1}{4}v^i_{jm}v^m_{kl}
\ee
That is tangent space can be equipped with composition laws (\ref{tang})
and the resulting binary-ternary algebra is called a tangent loop algebra.
Note, that if analytic loop is associative, i.e. is Lie algebra,
then ternary operation is equivalent to zero $ <\xi,\eta,\zeta>=0$
for all $\xi, \ \eta, \ \zeta \in g$. 
If analytic loop is binary associative loop (alternative loop) \cite{a3}
then the ternary operation is completely defined by the binary
operation $ \ <\xi,\eta,\zeta>=(1/6) J[\xi,\eta,\zeta] $.
In normal coordinates of arbitrary local analytic loop we obtain
$ x \circ y= x+y+(1/2)xy+... $ , 
where $ \ xy$ -- binary anti-symmetric  ($xx=0$ or $xy=-yx$ ) operation 
for elements of tangent loop algebra.
Let us prove that a loop commutator can be 
written by
\be
\label{form1}
 [x,y] \equiv (x \circ y) \circ (x \circ y)^{-1}= xy + ... 
\ee
where dots denote  combination of degree more than 3. Really
\[  x \circ y =  x+y+(1/2)xy+... \ , \qquad
 (x\circ y)^{-1} = -x-y-(1/2)xy-...\]
\[  [x,y]=  (x \circ y) \circ (x \circ y)^{-1}=x+y+ (1/2)xy - \]
\[ -x-y-(1/2)xy+(1/2)(x+y)(-x-y)+... = -(1/2)(xx-yy - 2yx) \ . \]
For anti-symmetric operation ( $xx=0$ and $xy=-yx$ ) 
we derive the formula (\ref{form1}).

{\bf 4.4. Tangent Algebra of Commutant Associative Loop.}

A {\bf commutator} of the elements $x,y \in G$ is an element
$(x \circ  y) \circ (y \circ x)^{-1}$.
Let us remind that  
{\bf commutant} $G^{(-)}$ of a loop $G$ is
a subloop, which is generated by commutators of this loop
$ (x \circ y)  \circ (y \circ x)^{-1} $ ,  where $x,y\in G$.

Let us introduce new definition \\
DEFINITION 7.
A {\bf commutant associative loop (Valya loop)} 
is a loop with inverse elements the commutant of which
(a set of commutators) is associative subloop (group). 

Analog of Lie theorem is formulated in the following form

THEOREM. 

a. A tangent algebra of local analytic commutant associative loop
(Valya loop) is a commutant Lie algebra (Valya algebra).

b. An arbitrary Valya (commutant Lie) algebra is a tangent algebra of 
some local analytic Valya (commutant associative) loop.

PROOF:

a. Let us consider an arbitrary element of a loop commutant $G^{(-)}$.
According to the definition an element of the commutant can be represented
in the form $z_1z_2...z_m$, where $z_i$ is a commutator of elements 
$x_i \ y_i$ of the loop $G$. If a set $G$ is a commutant associative loop, 
then the product $g_1(t_1)g_2(t_2)...g_m(t_m)$, where 
$g_i(t_i)$ are arbitrary one-parameter subgroups
with tangent vectors $z_i$, is not depends on brackets order
for all small $t_i$. Therefore, a subalgebra of tangent 
loop algebra, which is generated by elements $z_i$, is Lie algebra, 
that is the tangent algebra is commutant Lie algebra (Valya algebra).

b.  Let us introduce a normal coordinates.
Note, that one-parameter subgroups $g_i(t_i)$ 
generate a local associative subloop. It is easy to see that 
in normal coordinates we have Teylor power series,
which allows to derive a loop from an algebra.

Note that the proof of part (b) in general case is open question.


\section{  Realization of Commutant Lie Algebra as
Algebra of 1-Forms.}

{\bf 5.1. Poisson Algebra for Differential 1-Forms.} 

It is known that Poisson brackets can be defined for non-closed
differential 1-forms on symplectic manifold \cite{Maslov}. 
Poisson brackets for two 1-forms  
$ \alpha = a_k (z) d z^k  $ and $\beta =  b_k (z) d z^k $ 
on symplectic manifold $(M,\omega)$, 
is 1-form $(\alpha, \beta)$, defined by
\be
(\alpha, \beta) \ = \ d \Psi (\alpha, \beta) \ + \ \Psi (d \alpha, \beta) \ 
+ \ \Psi(\alpha, d \beta)  
\ee
\be
where \ \  \Psi(\alpha, \beta) \ = \ \omega (X_{\alpha}, X_{\beta}) 
\ = \ \Psi^{kl} a_k b_l \ ;  
\ee
and $ X_{\alpha}$ -- vector field, which corresponds to 1-form 
$\alpha$ by the rule:$ \ i(X_{\alpha}) \omega =  \alpha \ ; \ $
$ \omega  $ - closed ($d \omega = 0$) non-degenerate 2-form 
is called a symplectic form; $ \ \ i$ --  internal multiplication of
vector fields and differential forms \cite{Godbion}; 
$\Psi$ is cosymplectic structure and $ \Psi^{kl} $ 
is 2-tensor, which is a matrix
inverse to matrix of symplectic form and satisfies
the following conditions \cite{Maslov}:

a)  \ Skew-symmetry: $ \qquad \quad \   \Psi^{kl} =  \Psi^{lk} $

b) \ Zero Schouten brackets: 
\[  [ \Psi, \Psi]^{slk} =  \Psi^{sm} \partial_m \Psi^{lk} + 
\Psi^{lm} \partial_m \Psi^{ks} +  \Psi^{km}  \partial_m \Psi^{sl} = 0  \]

If this bilinear operation "Poisson bracket" is defined on the 
space of 1-forms  $ \Lambda^1 (M)$, then a manifold $M$ is called 
Poisson manifold, and the space $ \Lambda^1 (M)$ with this operation
-- Poisson algebra $P_1$. 
Poisson algebra $P_1$ is Lie algebra. It is cased by skew-symmetry  
$(\alpha, \beta) = - (\beta, \alpha)$ and Jacobi identity:
\[ J(\alpha,\beta,\gamma) = ((\alpha, \beta), \gamma) + 
((\beta , \gamma) \alpha) + ((\gamma , \alpha), \beta) \ = \ 0 \]

{\bf 5.2. Non-Closed Forms and Non-Hamiltonian Systems.}

It is known that a {\bf physical system} in classical mechanics 
is a vector field $X$ on the symplectic manifold $(M^{2n},\omega)$.

Non-Hamiltonian and dissipative properties of the system
are connected with properties of non-closed differential form
which can be derived from correspondent vector field.

Following \cite{Godbion} we shall call
{\bf Hamiltonian system} on the symplectic manifold 
$(M^{2n},\omega)$  a vector field
$X$ such that differential 1-form $i_{X} \omega$ is closed.

If the form $i_{X} \omega$ is exact form, then {\bf Hamiltonian} 
of the system $X$ is a function $H$ on $M^{2n}$, which
satisfies the following equation $i_{X} \omega=-dH$.

Let us introduce the following obvious definition: \\
DEFINITION 8.
{\bf Non-Hamiltonian (dissipative) system} on symplectic manifold
$(M^{2n},\omega)$ is a vector fields $X$ such that
the differential 1-form  $i_{X} \omega$ is non-closed.

{\bf 5.3. Generalized Poisson Algebra for Differential 1-Forms.} 

In order to describe non-Hamiltonian systems we suggest
\cite{Tar3} to generalize Poisson algebra $P_1$. 
Let us define a new operation on the space $ \Lambda^1 (M)$ . 
 
DEFINITION 9.
A {\bf generalized Poisson bracket} of two 1-forms $\alpha$ and $\beta$ is 
a closed 1-form $(\alpha, \beta)$, defined by
\be
\label{gpb}
[ \alpha, \beta ] \ = \ d \ \Psi (\alpha, \beta) \ =  
\ d \ \omega (X_{\alpha},X_{\beta}) 
\ee
It is easy to see that Jacobi identity for non-closed 1-forms
is not satisfied.
\[ J[\alpha,\beta,\gamma] = [[\alpha, \beta], \gamma] + 
[[\beta , \gamma] \alpha] + [[\gamma , \alpha], \beta] \ \not \equiv \ 0 \]
Therefore generalized Poisson algebra $P^{*}_1$ 
is non-Lie algebra. Note that Jacobi identity is satisfied for closed 1-forms.
So closed 1-forms define Lie algebra, which is Poisson algebra
$P_1$ of closed 1-forms. 
In the result, generalized Poisson algebra $P^*_1$ contains
a subalgebra which is Poisson algebra.
Generalized Poisson bracket of two non-closed 1-forms is
closed 1-form, so the subalgebra $P_1$ is an ideal of the algebra
$P_1^*$ and exact diagram exists:
$ 0 \rightarrow  P_1  \rightarrow P_1^*  \rightarrow  P_1^* / P_1 
  \rightarrow  0 $. 

Since generalized Poisson bracket is
a closed 1-form, the following proposition easy to prove

PROPOSITION.\\
A generalized Poisson algebra is a commutant Lie algebra (Valya algebra),
that is a generalized Poisson bracket (\ref{gpb}) satisfies
anti-symmetric identity $ [\alpha, \alpha] = 0 $ and 
soft Jacobi identity:
\[  J[[\alpha , \beta] ,[\gamma,\delta], [\mu,\nu]] =0 \]

Finally, the commutant Lie algebra, which is non-Lie algebra,
can be naturally defined in the space of differential 1-forms
$ \Lambda^1 (M)$ on symplectic manifold $M$. This algebra 
is generalization of Lie-Poisson algebra of closed 1-forms 
and contains this Lie algebra as subalgebra (ideal).

An obvious application of Valya algebra and Valya loop is a 
 
quantum description of non-Hamiltonian (dissipative) systems.
 
Let us note in the conclusion that non-Hamiltonian (dissipative)
quantum theory has a broad range of application, see the
references in \cite{Tartmf1,Tarictp}.
For example, consistent theory of bosonic string in affine-metric
curved space is non-Hamiltonian quantum theory \cite{Tarpl,Tartmf2,Tarmpl}.



\end{document}